\documentclass{emulateapj}
\usepackage{epstopdf}
\usepackage{placeins}
\usepackage{natbib}
\usepackage{verbatim}

\newcommand{\lya}{Ly-$\alpha$~}

\newcommand{\pcmsq}{cm$^{-2}$}
\newcommand{\cii}{C~{\sc ii}~}
\newcommand{\ciii}{C~{\sc iii}~}
\newcommand{\civ}{C~{\sc iv}~}
\newcommand{\cv}{C~{\sc v}~}

\newcommand{\oi}{O~{\sc i}~}

\newcommand{\siii}{Si~{\sc ii}~}

\newcommand{\siiv}{Si~{\sc iv}~}
\newcommand{\hi}{H~{\sc i}~}

\newcommand{\heii}{He~{\sc ii}~}

\newcommand{\heiinsp}{He~{\sc ii}}

\newcommand{\civnsp}{C~{\sc iv}}
\newcommand{\cvnsp}{C~{\sc v}}
\newcommand{\cvinsp}{C~{\sc vi}}

\newcommand{\oinsp}{O~{\sc i}}

\newcommand{\ciinsp}{C~{\sc ii}}
\newcommand{\siiinsp}{Si~{\sc ii}}
\newcommand{\siivnsp}{Si~{\sc iv}}
\newcommand{\feiinsp}{Fe~{\sc ii}}
\newcommand{\mgiinsp}{Mg~{\sc ii}}

\newcommand{\aliinsp}{Al~{\sc ii}}
\newcommand{\feii}{Fe~{\sc ii}~}
\newcommand{\alii}{Al~{\sc ii}~}

\newcommand{\mgii}{Mg~{\sc ii}~}
\newcommand{\mgi}{Mg~{\sc i}~}

\newcommand{\mciv}{{\rm C \; \mbox{\tiny IV}}}

\newcommand{\kms}{km s$^{-1}$}

\begin{document}

\title{Constraints on the Universal \civ Mass Density at $z\sim 6$\\
  from Early IR Spectra Obtained with the Magellan FIRE Spectrograph\altaffilmark{1}} 

\author{Robert A. Simcoe\altaffilmark{2,8}}
\author{Kathy L. Cooksey\altaffilmark{2,9}}
\author{Michael Matejek\altaffilmark{2}}
\author{Adam J. Burgasser\altaffilmark{2,3,10}}
\author{John Bochanski\altaffilmark{2,4}}
\author{Elizabeth Lovegrove\altaffilmark{2,5}}
\author{Rebecca A. Bernstein\altaffilmark{5}}
\author{Judith L. Pipher\altaffilmark{6}}
\author{William J. Forrest\altaffilmark{6}}
\author{Craig McMurtry\altaffilmark{6}}
\author{Xiaohui Fan\altaffilmark{7}}
\author{John O'Meara\altaffilmark{11}}

\altaffiltext{1}{This paper includes data gathered with the 6.5 meter
  Magellan Telescopes located at Las Campanas Observatory, Chile.}
\altaffiltext{2}{MIT-Kavli Center for Astrophysics and Space Research}
\altaffiltext{3}{Center for Astrophysics and Space Science, University
  of California, San Diego, La Jolla, CA, 92093, USA}
\altaffiltext{4}{Pennsylvania State University, Department of Astronomy
  and Astrophysics, 525 Davey Lab, University Park, PA 16802, USA}
\altaffiltext{5}{University of California, Santa Cruz}
\altaffiltext{6}{University of Rochester} \altaffiltext{7}{University
  of Arizona} \altaffiltext{8}{Sloan Foundation Research Fellow}
\altaffiltext{9}{NSF Fellow} \altaffiltext{10}{Hellman Fellow}
\altaffiltext{11}{St. Michael's College}

\begin{abstract}
We present a new determination of the intergalactic \civ mass density
at $4.3 < z < 6.3$.  Our constraints are derived from high
signal-to-noise spectra of seven quasars at $z>5.8$ obtained with the
newly commissioned FIRE spectrograph on the Magellan Baade telescope,
coupled with six observations of northern objects taken from the
literature.  We confirm the presence of a downturn in the \civ
abundance at $\langle z\rangle=5.66$ by a factor of $4.1$ relative to
its value at $\langle z\rangle=4.96$, as measured in the same
sightlines.  In the FIRE sample, a strong system previously reported
in the literature as \civ at $z=5.82$ is re-identified as \mgii at
$z=2.78$, leading to a substantial downward revision in $\Omega_\mciv$
for these prior studies.  Additionally we confirm the presence of at
least two systems with low-ionization \ciinsp, \siiinsp, and \oi
absorption but relatively weak signal from \civnsp.  The latter
systems systems may be of interest if the downward trend in
$\Omega_\mciv$ at high redshift is driven in part by ionization
effects.
\end{abstract}

\section{Introduction}\label{sec:intro}

In March 2010, our group commissioned the Folded-Port Infrared
Echellette (FIRE), a new IR echelle spectrograph for the 6.5 meter
Magellan Baade Telescope.  FIRE delivers $R=6000$ spectra with
uninterrupted coverage from $0.82$ to $2.5 ~\mu$m for a
$0.6^{\prime\prime}$ slit \citep{FIRE_SPIE2,FIRE_PASP}.  One of the
major science drivers for its construction was exploration of the
high-redshift intergalactic medium (IGM) via absorption line
spectroscopy.  Here, we report initial results on the search for
high-redshift \civ based on commissioning observations taken during
the first year of FIRE's operation.

The existence of heavy elements in the intergalactic medium (IGM) has
been known since the earliest detections of \civ associated with
$z\sim 3$ \lya forest absorbers in high resolution optical spectra
\citep{cowie_civ_1,songaila_civ,cowie_civ_nature}.  Soon after this
discovery, intergalactic \civ lines were detected even in the spectra
of quasars with $z>4$, establishing a lower limit on the amount of
metal production that must have occurred in the early universe
\citep{songaila_omegaz}.  A surprising result of these studies was the
relatively constant comoving number density of \civ atoms over a wide
range in redshift, where this density is smoothed over very large
scales.  More recent UV observations at low redshift
\citep{cooksey_civ, danforthshull} point toward a slow rise in the
\civ density---expressed as $\Omega_\mciv$, the \civ contribution to
closure---increasing by a factor of three to four between $z\sim 2$ and the
present day.

The discovery of luminous quasar populations at $z>6$
\citep{fan_z6qsos_paper4,willott_cfqs, mortlock_ukidss} enables
studies of the \civ density at higher redshift, but above $z\sim 5.5$
observations become challenging as the \civ transition moves into the
$Y$ and $J$ bands.  For this reason, the earliest pilot studies in
this region focused on very small (i.e. $N=2$) samples of quasar
sightlines.  These pilot studies yielded similar values of
$\Omega_\mciv$ as at lower $z$, but with few actual line detections
they were strongly affected by small number statistics
\citep{simcoe_z6,ryan_weber_1,becker_civ}.  More recently
\citet{ryan_weber_civ} compiled a larger set of IR observations for 9
QSOS, with VLT/ISAAC and Keck/NIRSPEC.  Their dataset contains several
sightlines with no \civ detections, which points to a decrease in
$\Omega_\mciv$ at $z>5.5$.

During FIRE's first year of operation we have obtained spectra of
eleven $z>5.5$ quasars, of which eight have suitable signal-to-noise
ratio (SNR) for IGM abundance measurements.  This paper presents a
first analysis of seven of these spectra, focusing on the search for
\civ doublets in particular.  We combine our results with data from
the literature to derive a new value for $\Omega_\mciv$ based on 13
unique sightlines.  Where relevant throughout the remainder of the
paper, we adopt a flat cosmology with $\Omega_M=0.3$,
$\Omega_\Lambda=0.7$, and $H_0=71\,{\rm km}\,{\rm s}^{-1}\,{\rm Mpc}$.

\section{Observations and Data Reductions}\label{sec:observations}

During the period from 31 March 2010 to 3 April 2011, we obtained FIRE
spectra of 11 QSOs with $z_{em}>5.5$, of which eight had high enough
SNR to study IGM absorption.  Table 1 lists details of the
observations for the seven high-redshift targets included in the
present analysis.  All spectra were obtained with a
$0.6^{\prime\prime}$ slit, in seeing conditions ranging from
$0.35$--$0.9\arcsec$, for a measured resolution of $R=6000$, or 50 \kms.
Individual integration times are listed in the table, with typical
values of 4-5 hours.  Full integrations were broken up into 15- to
20-minute intervals between which the object was moved on the slit.

We reduced the data using a custom-developed IDL pipeline, evolved
from the MASE suite used for optical echelle reduction \citep{MASE}.
There were several important modifications implemented for IR work.
The most important of these was for wavelength calibration, which is
principally solved using the terrestrial OH lines imprinted upon
science spectra.  Unlike most IR spectral packages which perform
pairwise subtraction of A/B slit positions for sky subtraction, the
FIRE pipeline instead calculates a direct B-spline model of the sky
for each exposure using the techniques of \citet{kelson_bspline}.
This yields Poisson-limited residuals in the sky-subtracted frame, for
a theoretical improvement of $\sqrt{2}$ in SNR relative to pairwise
subtraction.  This pipeline is being released to the community as part
of the instrument package; a full description of its functionality and
the efficacy of the sky subtraction will be provided in a separate
forthcoming paper.

\begin{figure*}[t]
\plotone{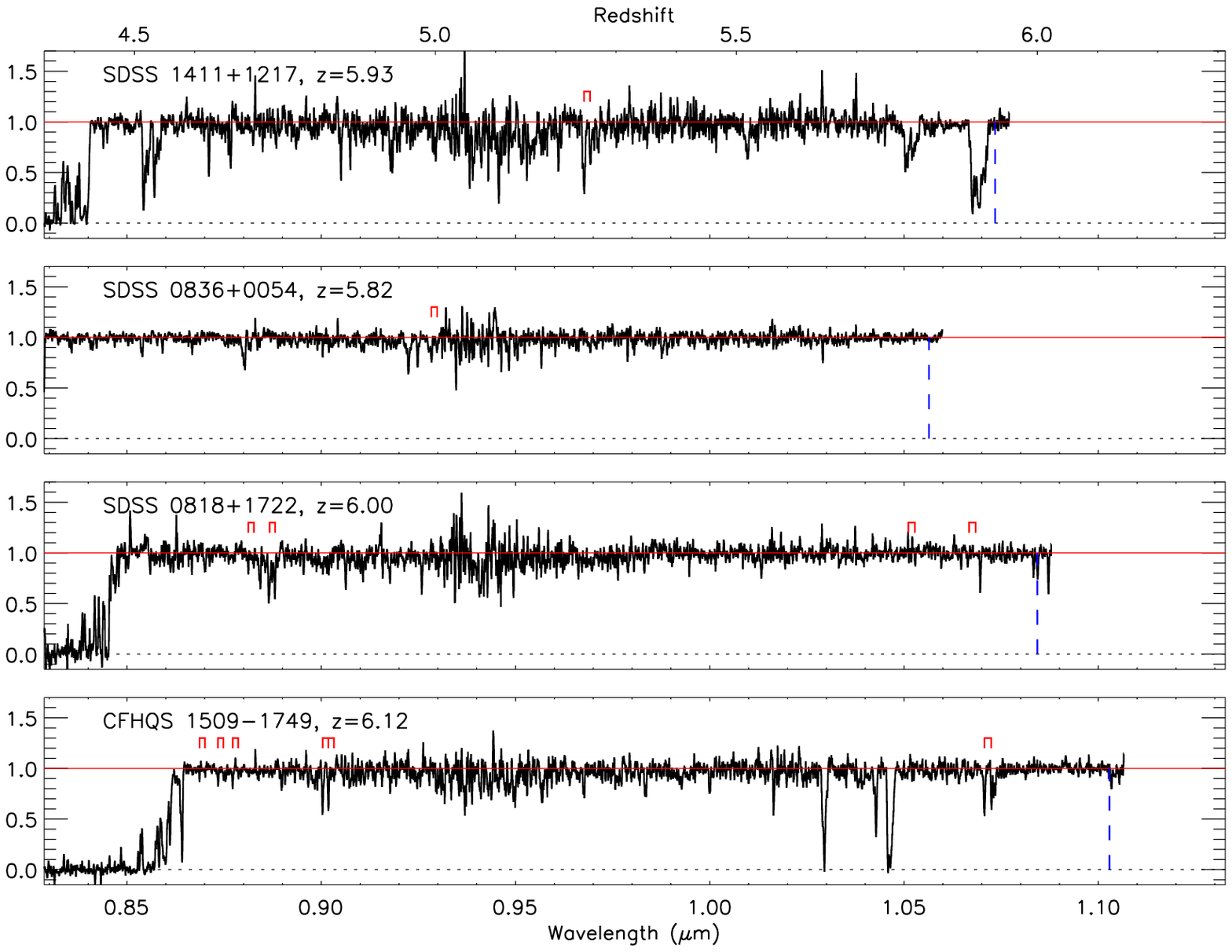}
\plotone{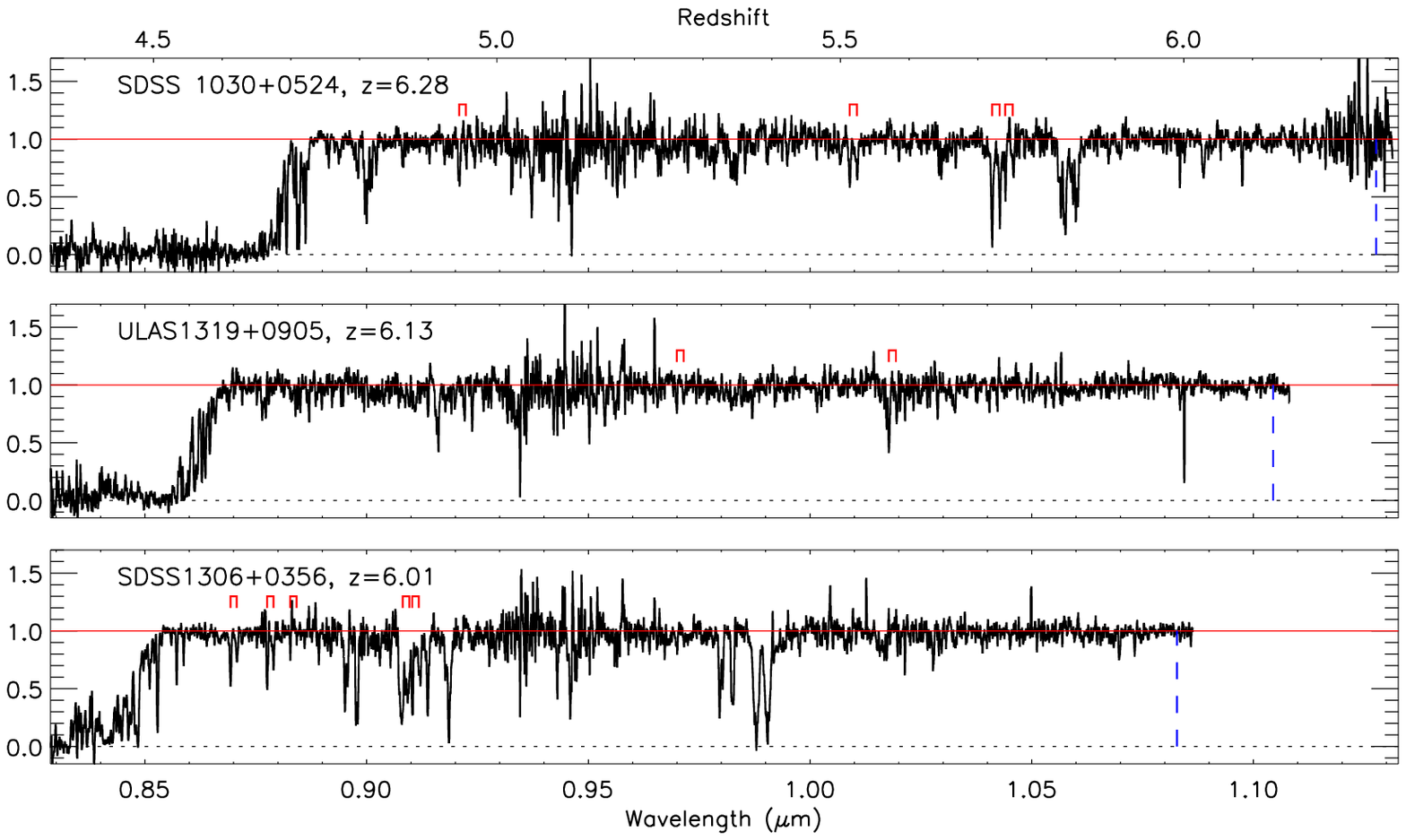}
\caption{Normalized spectra of the seven sample objects, ranging from
  the minimum search redshift ($z=4.35$ in SDSS0836+0054) to the
  maximum $z=6.3$ (in SDSS1030+0524).  The locations of \civ doublet
  in Table 2 are shown with red marks, and the \civ wavelength at the QSO
  emission redshift is shown as a vertical dashed line.  The weaker \civ
  systems that are difficult to see in this plot are better displayed
  in Figure \ref{fig:systems}.}
\label{fig:sightlines}
\end{figure*}

During observations, we obtained contemporaneous spectra of A0V stars
for correction of telluric absorption features using the method of
\citet{vacca_telluric}.  We used the {\tt xtellcor} procedure released
with the {\tt spextool} pipeline \citep{spextool} to perform telluric
correction and relative flux calibration, finally combining the
corrected echelle orders from all exposures into a single, 1D spectrum
for analysis.

For each sample target, we obtained additional far-optical spectra
using the Magellan Echellette spectrometer \citep[MagE, ][]{Marshall},
for the purpose of augmenting SNR in the region blueward of 9000\AA.
These spectra had typical integration times of $\sim 1$ hour, and
$R=5600$, a close match to FIRE.  The data were processed in similar
fashion as described above.  Rather than performing detailed telluric
corrections to the MagE data, we instead masked out regions
susceptible to telluric contamination when combining with the FIRE
spectra, only including the MagE contribution in telluric-free areas.
The MagE spectra were smoothed to match the resolution of FIRE and the
spectra from the two instruments were combined with inverse variance
weighting, to create a single composite for analysis.

\section{Analysis}

\subsection{Line Identification}\label{sec:ids}

Before searching the data for \civ doublets, we fit a smooth continuum
estimate to each quasar spectrum using a slowly-varying cubic spline.
Figure 1 displays the \civ region for each of the continuum-normalized
spectra.  These regions were then searched both by hand and also using
automated software to identify \civ doublets.

The automated search algorithm followed a two step process.  First,
the inverse spectrum ($1-f/f_0$, where $f$ is the flux and $f_0$ the
continuum) was smoothed with a Gaussian kernel having full-width at
half-maximum (FWHM) of one spectral resolution element---i.e. 50 \kms
~or 4 pixels.  A peak finding procedure was run on the convolved
spectrum to identify all deviations of $\ge 2.5\sigma$, which after
convolution is equivalent to the SNR per resolution element.

\begin{deluxetable}{lccccc}
\tablecaption{FIRE Observations}
\tablehead{
\colhead{Object} &
\colhead{$z$} &
\colhead{$J$\tablenotemark{1}} &
\colhead{Exptime (s)} &
\colhead{$\Delta z$} &
\colhead{$\Delta X$}}
\startdata
SDSS 0818+1722  & 6.00 & 18.5 & 9,000  & 4.50-6.00 & 5.61 \\
SDSS 0836+0054  & 5.82 & 17.9 & 10,187 & 4.35-5.79 & 4.52 \\
SDSS 1030+0524  & 6.28 & 18.9 & 14,400 & 4.75-6.29 & 6.82 \\
SDSS 1306+0356  & 6.01 & 18.8 & 15,682 & 4.52-5.97 & 6.46 \\
ULAS 1319+0905  & 6.13 & 18.9 & 19,275 & 4.62-6.08 & 6.56 \\
SDSS 1411+1217  & 5.93 & 18.9 & 15,300 & 4.55-5.93 & 5.28 \\
CFHQS 1509--1749 & 6.12 & 18.9 & 17,100 & 4.60-6.11 & 6.03 
\tablenotetext{1}{Vega magnitudes}
\enddata
\end{deluxetable}

\begin{figure*}[t]
\epsscale{1.1}
\plottwo{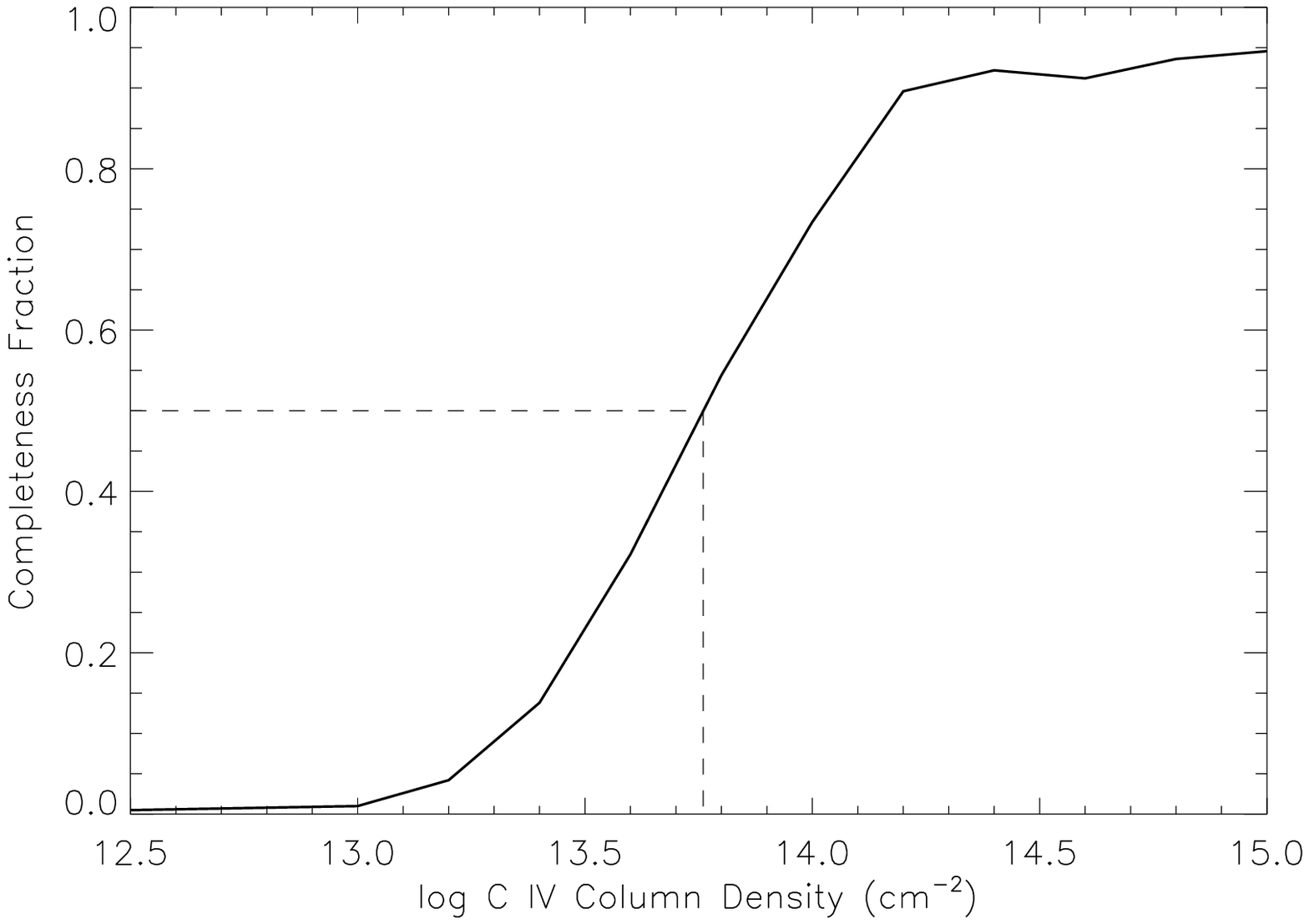}{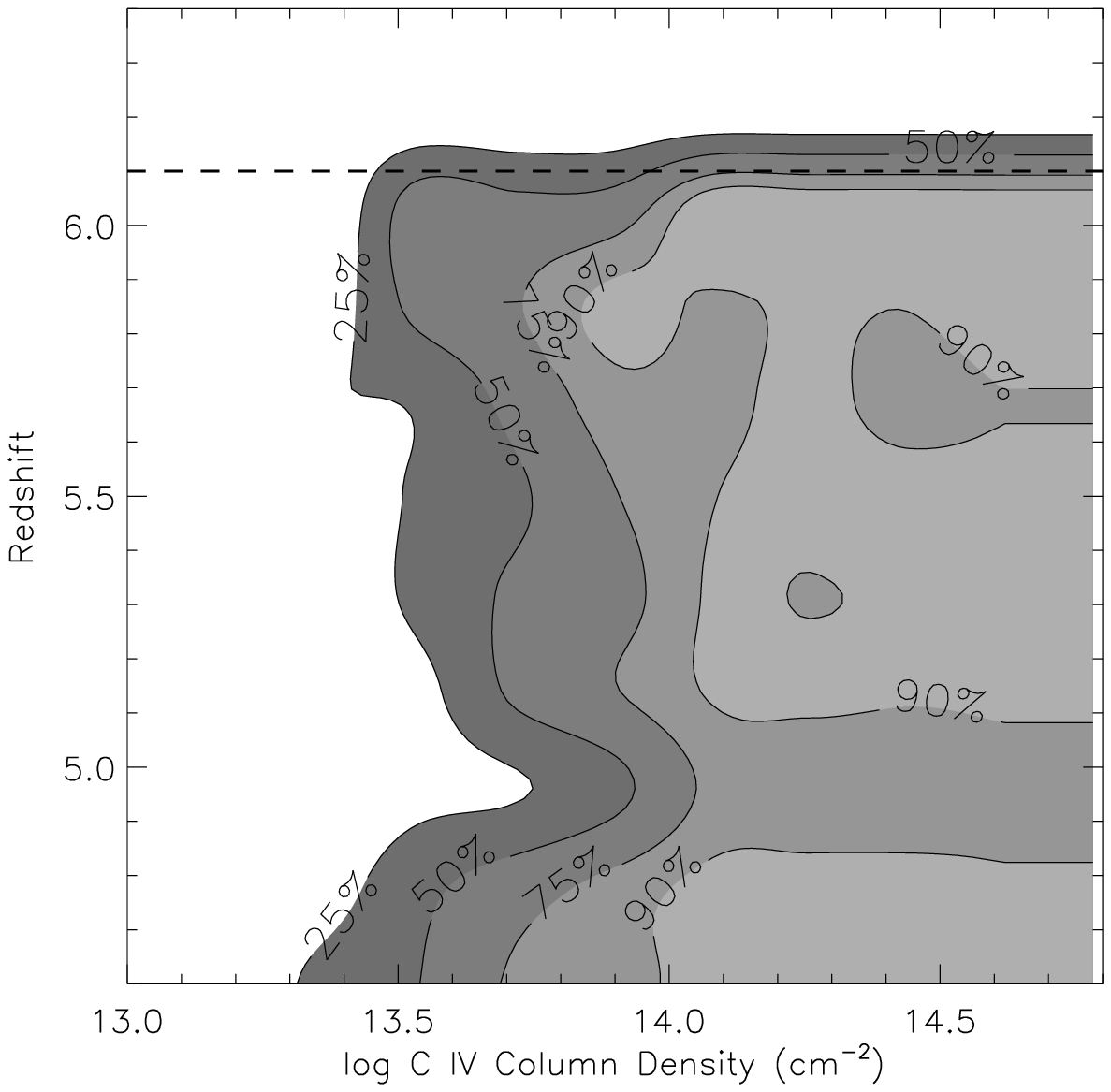}
\caption{{\em Left:} The combined \civ sample completeness averaged
  over the full path of the survey.  Dotted lines show the 50\%
  completeness level, which occurs around $N_\mciv=10^{13.75}$\pcmsq.
  {\em Right:} The distribution of completeness with \civ column
  density and redshift.  Some redshift ranges---particularly near
  $z=5.0$---have lower completeness because of higher noise levels in
  the data.  This dependence is taken into account when correcting
  each detected line up for incompleteness when determining
  $\Omega_\mciv$.} \epsscale{1.2}
\label{fig:completeness}
\end{figure*}

This list of single-line identifications was then parsed to identify
pairs at the appropriate velocity spacing for \civ doublets.  We
further required that the ratio of the absorption amplitudes between
the 1548\AA ~and 1551\AA ~lines be consistent with the ratio in
oscillator strengths, i.e. between 2.0 (for an unsaturated doublet)
and 1.05 (completely saturated).  Each candidate whose doublet ratio
fell within $\pm 1\sigma$ of this range was inspected visually to
reject obvious anomalies, for example from poor sky subtraction near
OH sky lines.  We also verified that candidate \civ pairs were not
\siiinsp/\oi doublets at higher redshift, by inspecting the
corresponding \civ and \mgii wavelengths, as well as \siii
$\lambda1526$\AA ~for each system that could be mis-identified in this
way.

\subsection{Completeness Estimates}\label{sec:completeness}

At $z\lesssim 4$, the \civ column density distribution has a power law
form extending at least to $N_\mciv = 10^{12.5\!-\!13.0}$ and possibly
lower, as measured from high SNR, high resolution optical spectra
\citep{ellison_civ,songaila_omegaz, songaila_new_civ}.  Our high
redshift sample is incomplete at these lower column densities, because
of our $\sim 10\times$ lower resolution (e.g. compared to HIRES) and
$\sim 5\times$ lower SNR when compared with the best optical QSO
spectra in the literature.

To quantify this incompleteness, we ran a simple set of Monte Carlo
tests where a simulated \civ signal was injected into our actual QSO
spectra.  Our line finding algorithm was then run to determine the
recovery fraction as a function of redshift, column density, and line
width $b$.

The intrinsic properties of $z\sim 6$ \civ lines are not yet
well-characterized because the largest existing surveys are likely
limited by instrumental resolution rather than intrinsic line
widths\footnote{The NIRSPEC spectra of \citet{becker_civ} with
  resolution 23 \kms ~may sample the high end tail of the $b$
  distribution at lower SNR, but the present study and others in the
  literature are a factor of $\sim 2$ lower resolution.}.  At lower
redshift, typical $b$ parameters take values in the range $b\sim 8-20$
\kms, with a slight preference for larger $b$ at higher column
densities.  For the completeness tests, we took a forward-modeling
approach by drawing $b$ parameters at random from a distribution
matched to HIRES \civ fits at $z\sim 2-3$
\citep{rauch_civ_kinematics}.  This distribution is Gaussian in shape,
and centered at 10 \kms ~for $N_\mciv=10^{13}$.  It has a cutoff below
5 \kms and a steady rise in the mean $b$ to 15 \kms at
$N_\mciv=10^{14}$ but again cuts off above $b\sim 25$ \kms.

Over a grid of column density, we drew redshifts at random, with a
distribution matched to the pathlength-weighted redshift distribution
of the survey sightlines.  We then calculated Voigt profiles directly
for each artificial line, convolved these with the instrumental
response function, and inserted them to the survey spectra, with one
artificial system, per spectrum, per trial.

We then re-ran the line search algorithm described above on each trial
spectrum and examined the output to see if the line was recovered.
Each successful recovery was inspected by eye to ensure that our
completeness was not being over-estimated due to chance correlations
in noise being picked up as ``detections'' for intrinsically weak
lines.  This procedure was repeated 5000 times, with 500 trials at
each point in a grid of $N_\mciv$ between $10^{13}$ and $10^{15}$
\pcmsq.

\begin{figure*}[t]
\epsscale{1.0}
\plotone{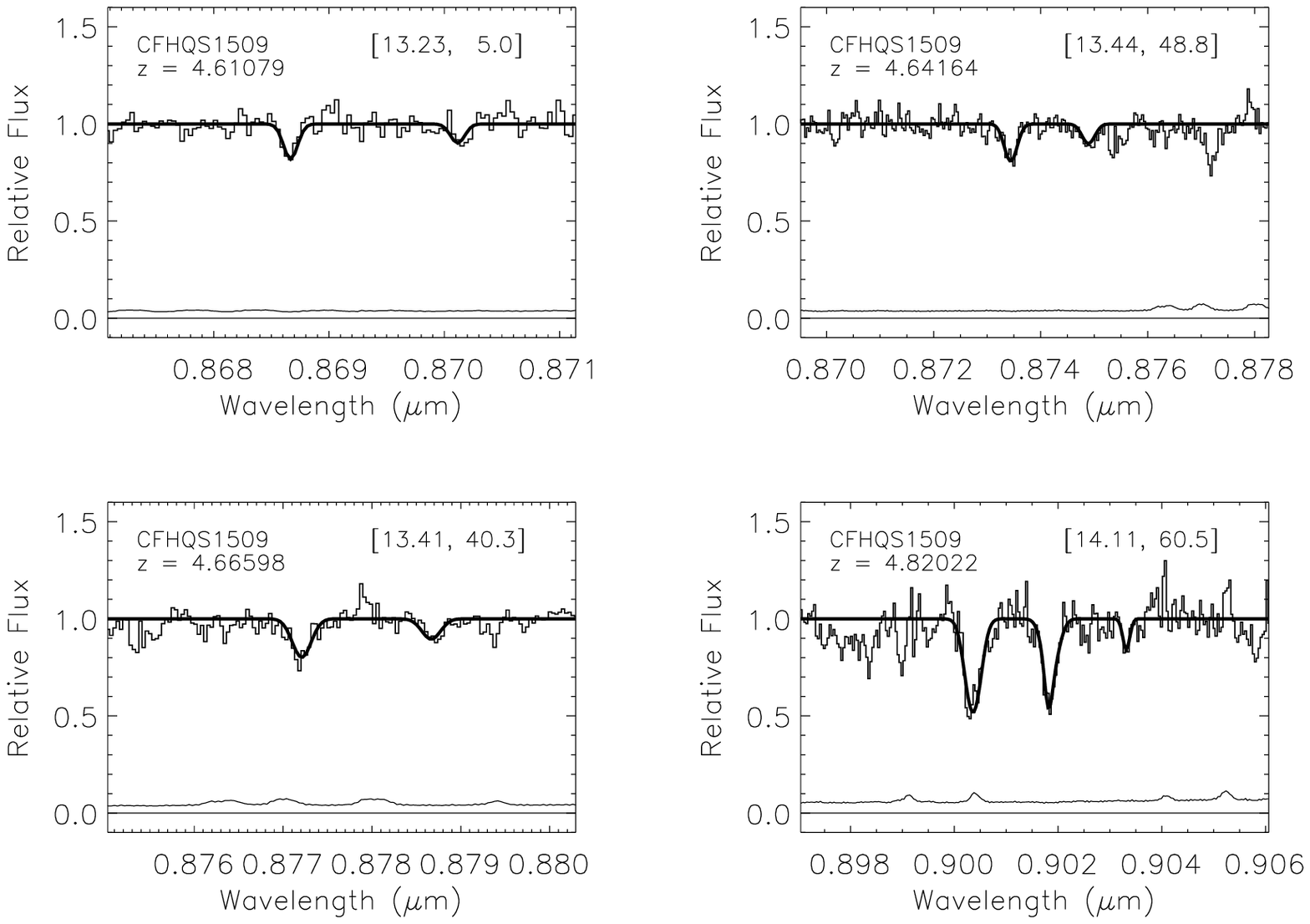}
\plotone{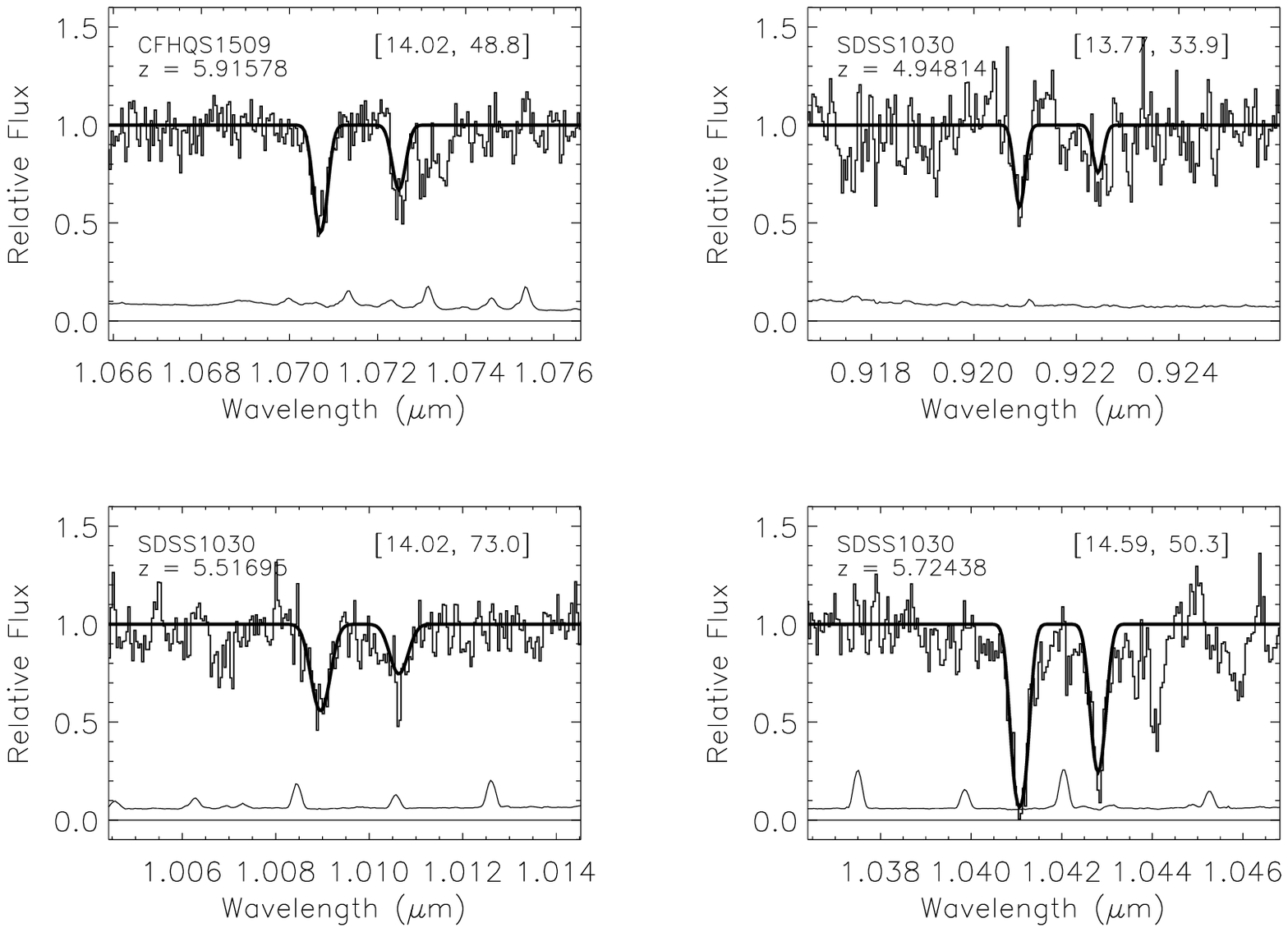}
\caption{Plots of individual \civ systems listed in Table 2, in units
  of continuum normalized flux.  The $1\sigma$ error array is shown at
  bottom, and a best-fit Voigt profile is shown with a solid smooth
  line.  The total $\log(N_\mciv)$ and $b$ is shown for each system in
  square brackets.}  \epsscale{1.0}
\label{fig:systems}
\end{figure*}

Figure \ref{fig:completeness} shows the total completeness
distribution of the survey; the left panel includes all $z$ while the
right panel shows the completeness map by both redshift and $N_\mciv$.
Broadly speaking, we recover most lines ($>80\%$) above $N_\mciv =
10^{14}$ and relatively few ($<10\%$) below $N_\mciv = 10^{13.2}$,
with the 50\% level occurring near $10^{13.7}$.  Thus our survey
sits between the parameter space explored by \citet{becker_civ}, who
are complete to lower $N_\mciv$ over a significantly smaller
pathlength, and \citet{ryan_weber_civ} who have a comparable
pathlength but slightly lower completeness.
 
From the right panel of Figure \ref{fig:completeness} it is also
apparent that the completeness varies with search redshift, with our
areas of highest sensitivity occurring at $z<4.8$ and $z>5.2$.  The
high completeness at $z>5.7$ reflects the rising sensitivity of FIRE
toward these wavelengths, while at $z<4.8$ the MagE spectra contribute
significantly to the SNR.  The relatively low completeness at $z\sim
5$ results from decreased flux and/or imperfect correction in the
vicinity of telluric absorption bands, which can be seen as increased
noise around $\lambda=0.94\mu$m in Figure \ref{fig:sightlines}.

In statistical calculations below, we correct for incompleteness by
scaling each detected line by the inverse of its estimated
completeness fraction.  It is worth noting that the \civ lines
reported in the literature to date nearly all have $N_\mciv > 10^{14}$
~\pcmsq; we should detect the large majority of these over our entire
survey volume, so the correction to $\Omega_\mciv$ from this source of
incompleteness is only $\sim 25\%$ or 0.1 dex - smaller than the
Poisson errors from the limited number of systems in the sample.  At
smaller column density the correction is more severe, but these
systems make a smaller fractional contribution to the total value of
$\Omega_\mciv$.

\section{\civ Sample and Measurements}\label{sec:measurements}

Using the criteria described above, we identified 19 unique \civ
systems ranging in redshift from $z=4.61$ to $z=5.92$ in the six
survey sightlines.  Properties of the individual systems are listed in
Table 2, and plots of the individual absorbers are shown in Figure
\ref{fig:systems}.  We excluded regions within $3000$ \kms ~of the QSO
emission redshift from the search; however, no lines were excluded by
this criterion, except for the one BAL absorber in SDSS1411.  The
Table contains two additional systems in SDSS0818+1722 at $z=5.789$
and $z=5.876$; these systems were not flagged as \civ in our data but
have been reported previously in the literature and are discussed in
Section \ref{sec:ISAAC_0818}.

We characterized the absorption strength for each system using a
variety of different methods and to test the robustness of our column
density determinations.  At a resolution of 50 \kms ~we almost
certainly fail to resolve the detailed velocity structure seen in \civ
systems at lower redshift, and we must also be attuned to the
possibility of unresolved saturated components.

For each system, we list the rest-frame equivalent width ($W_r$) of
both transitions of the \civ doublet, determined by direct summation
of the continuum-normalized spectral pixels.  The sums were performed
over a region 50 \kms above and below where the normalized profile
returns to unity.  We then list a \civ column density as determined
using the ``apparent optical depth'' method of \citet{AODM}, and
finally the column density and $b$ parameter estimates from direct
Voigt profile fitting using the {\tt vpfit} package.

Inspection of the Voigt profile Doppler parameters in Table 2 shows a
high average $b$, similar to what was described in
\citet{ryan_weber_civ}.  This width is more likely to represent the
full velocity spread of a suite of narrow components, rather than a
large, single component.  However, in such cases the {\em total}
column density of the system is represented surprisingly well in a
profile fit, since the ratio of the absorption depths reflects the
degree of saturation (i.e. the column density for a single component)
and the $b$ parameter then scales with the number of individual
sub-clumps in the system.

\subsection{SDSS0818+1722, $z_{em}=6.00$}\label{sec:SDSS0818}

We detect two discrete \civ systems toward this QSO, which had the
shortest total integration of our sample.  One of our detections, at
$z=4.7266$, is very robust and appears to contain at least two
subcomponents.  There is some hint of further substructure in the
profile but the SNR of our data did not warrant a more aggressive
fit.  The second system we detect is at $z=4.6919$ and is very weak
at $N_\mciv=10^{13.2\!-\!13.4}$.  

This sightline has also been observed with ISAAC
\citep{ryan_weber_civ}, NIRSPEC \citep{becker_civ} and XShooter
\citep{dodorico}.  From these studies the system at $z=4.7266$ is
confirmed but the system at $4.6919$ is not.  In addition,
\citet{ryan_weber_civ} and \citet{dodorico} report tentative \civ
detections at $z=5.7892$ and $z=5.8769$, respectively.  These systems
are discussed further in Section \ref{sec:ISAAC_0818}.  The $z=5.7892$
system appears to suffer from sky subtraction systematics in the FIRE
spectrum, and may be consistent with \civ absorption.  The $z=5.8769$
is detected in \civ 1548\AA, but the 1550\AA ~component of the doublet
narrowly misses our $2.5\sigma$ threshold.  However we confirm the
clear presence of low-ionization gas associated with both these
systems.  Since they are not formally detected in the FIRE spectrum,
we do not include these systems in our statistical calculations,
although they would represent only a 0.08 dex perturbation on the
final mass density calculation.  \citet{dodorico} report a very large
number of additional systems (13, at $z=4.498$, 4.508, 4.523, 4.552,
4.577, 4.620, 4.732, 4.877, 4.942, 5.076, 5.308, 5.322, and 5.344)
which we do not detect in our 9000s spectrum.  The systems at
$z\lesssim 5$ are detected in neither our MagE or FIRE spectra, and so
must be quite weak.  The higher redshift examples may only be revealed
at the higher SNR revealed by D'Odorico's 24,000s spectrum.

\subsection{CFHQS1509--1749}\label{sec:CFQS1509}

We detect 5 unique systems in our FIRE spectrum of CFHQS1509--1749.
Four of these are quite weak at $z<5$, while one at $z=4.820$ is
significant.  The systems at $z=4.6108, 4.6416,$ and $4.6659$ were
reported by \citet{dodorico}, but the system at $z=4.820$ was not.  A
second weak component was required to obtain a good fit on the
$z=4.8202$ system to avoid reversing the doublet ratio.

We detect an additional system not previously reported at $z=5.9158$;
this is the highest redshift \civ system in our sample, and to our
knowledge the highest redshift absorption system currently known.  The
\civ system listed at $z=4.820$ is closely matched in wavelength to
the expected location of the \oinsp/\siii pair, but this alternate
identification is ruled out by a lack of \siii $\lambda1526$\AA.

\subsection{SDSS1030+0524}\label{sec:SDSS1030}

We detect three distinct systems in SDSS1030+0524, and retract from
this sightline a fourth system that has been discussed in the
literature.  The first system is at $z=5.7244$ and was a strong
detection reported by \citet{ryan_weber_1} and a marginal detection in
\citet{simcoe_z6}.  In addition, we find new systems at $z=4.9481$ and
$z=5.5169$.  The $z=4.9481$ system is confirmed by the presence of
\siii 1526\AA, \alii 1670\AA, and the \mgii doublet, while the
$z=5.5169$ system is confirmed by \siii 1526\AA, \feii (multiple
lines), and \alii 1670\AA (\mgii falls between the $H$ and $K$ bands).
We detect an additional \civ doublet at $z=5.744$, in the
vicinity of the known system at $z=5.72$.  However the strong
absorption system at $1.057\mu$m---previously identified as \civnsp---is
shown in the FIRE spectrum to be a remarkable low redshift interloping
\mgii system (discussed further in Section \ref{sec:SDSS1030_mgii})

\subsection{SDSS0836+0054}\label{sec:SDSS0836}

This bright, lower redshift QSO has been observed by the groups noted
above, with no \civ detections.  Despite excellent SNR, we also detect
no \civ absorption at $z>5$, although we do detect a new system at
$z=4.996$, with possible \siiv seen at the $2\sigma$ level.  At $z>5$,
this sightline is roughly 0.2 dex more sensitive than the survey
average: its 50\% completeness falls at $\log(N_\mciv)=13.55$, and is
80\% complete at $\log(N_\mciv)=13.8$.

\setcounter{figure}{2}
\begin{figure*}[t]
\epsscale{1.0}
\plotone{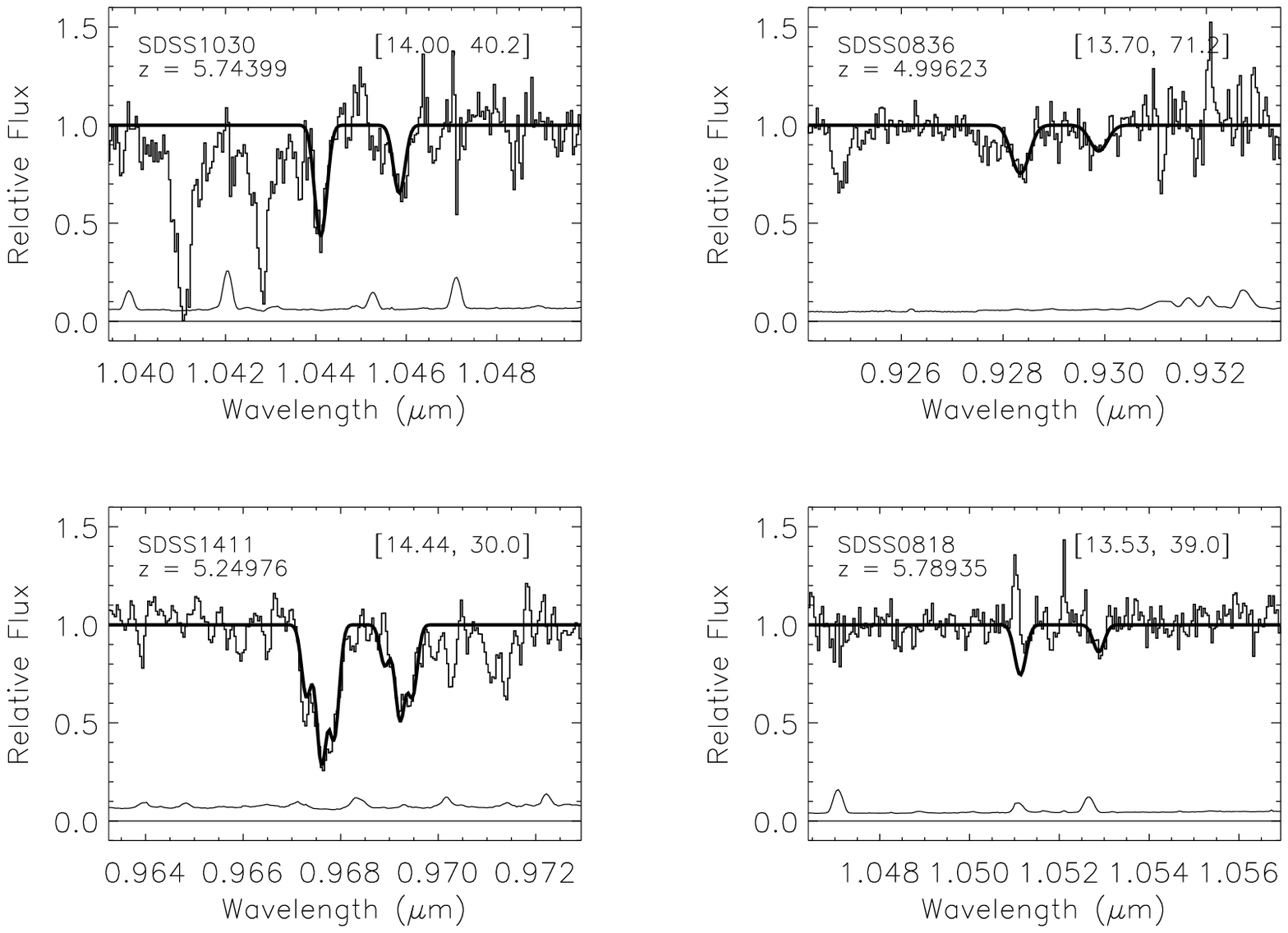}
\caption{{\em Continued}}
\epsscale{1.0}
\end{figure*}
\setcounter{figure}{2}
\begin{figure*}[h]
\epsscale{1.0}
\plotone{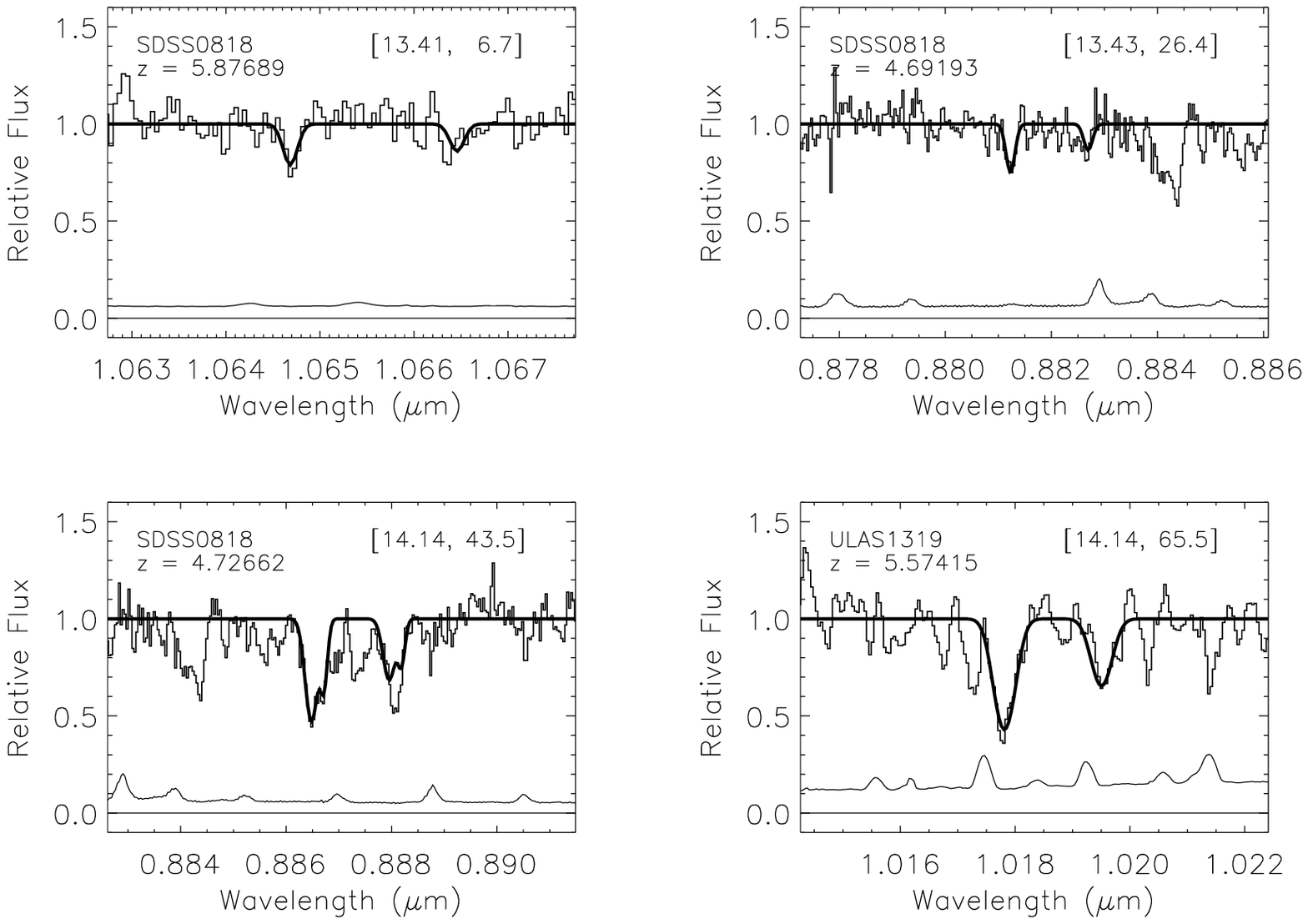}
\plotone{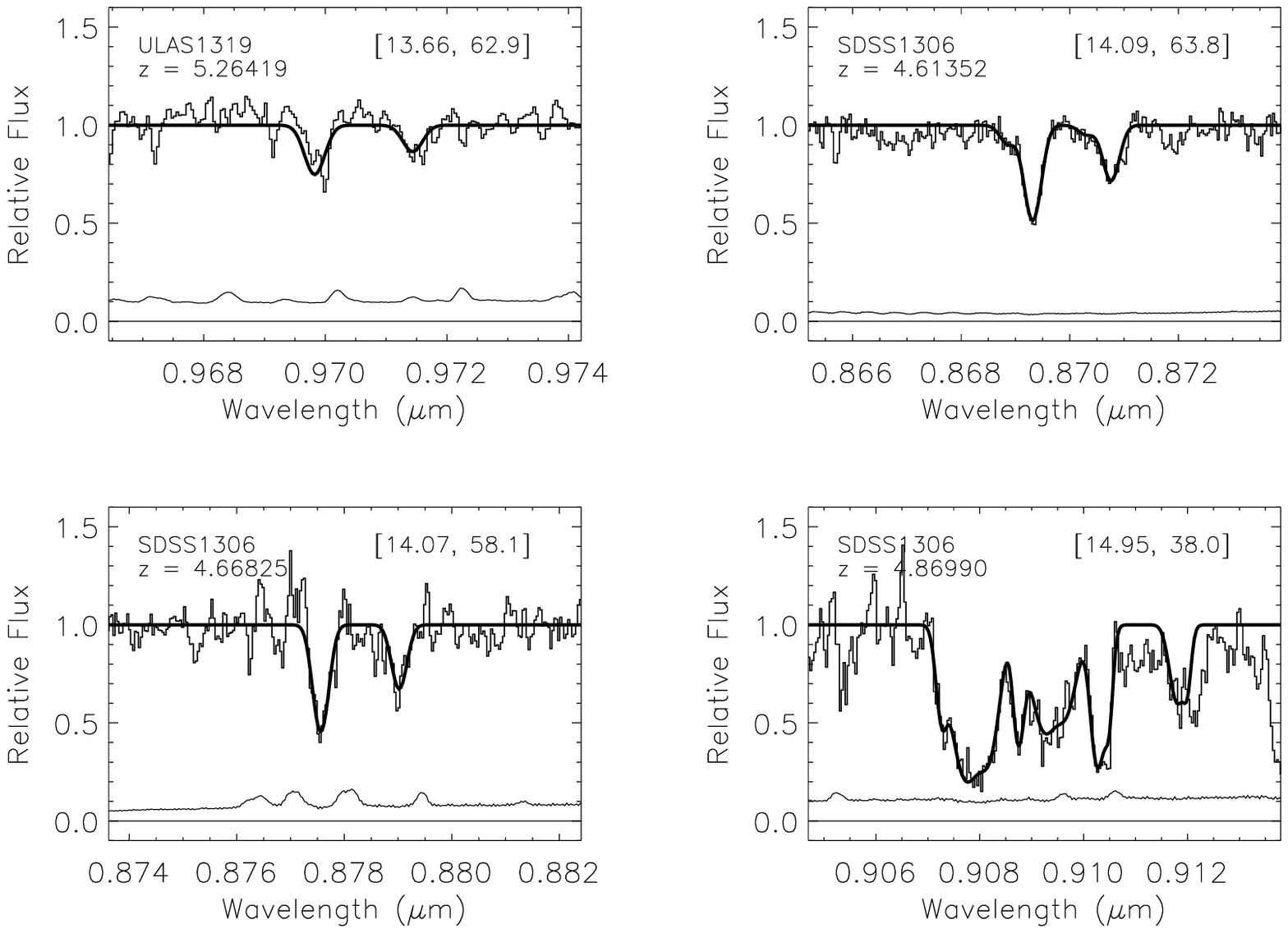}
\caption{{\em Continued}}
\epsscale{1.0}
\end{figure*}

\subsection{SDSS1411+1217}\label{sec:SDSS1411}

This spectrum had among the lowest SNR of our sample given its ratio
of emission line to continuum flux.  However we detected one very
strong \civ system at $z=5.2498$ which shows evidence of velocity
structure in its profile.  This system is also seen in \mgii and
\siivnsp; \cii and \siii are not detected but also not strongly ruled out
by the data.  \feii falls in a high SNR region of the spectrum but was
not detected.

There is a high redshift absorber that appears to be \civ near
$z=5.786$, but this line has extremely extended absorption as would be
characteristic of a BAL at the redshift of the background QSO, so we
do not include it in the IGM sample.

\subsection{ULAS1319+0950}

This sightline has not previously been observed for \civ absorption.
We detect a significant \civ system at $z=5.573$, which is accompanied
by strong \siiv, possible \cii, and \feii.  A second, very marginal
detection is seen at $z=5.2646$ with no other transitions.  More
sensitive data are needed to confirm the identification of this system.

\subsection{SDSS1306+0305}

Three \civ systems are detected in the lower-redshift region of this
spectrum, including a very strong complex at $z=4.870$ that also
displays \siiinsp, \feiinsp, \aliinsp, \mgiinsp, and possible \mgi.
This system was also seen in the optical spectra of \citet{becker_z6}.
Significant systems are also detected at $z=4.613$ and $4.686$; a weak
system is possibly present at $z=4.702$ but is only detected at
$2\sigma$ significance and so is not included in the sample.  Similar
to previous studies of this sightline, we find no statistically
significant \civ absorption at $z>5$.  The two strong absorption lines
at 9900\AA ~are from an interloping \mgii system at $z=2.53$.

\subsection{Special Cases}\label{sec:noteworthty}

\subsubsection{The $z=5.7899$ absorber in SDSS0818}\label{sec:ISAAC_0818}

\begin{figure}[t]
\epsscale{1.2}
\plotone{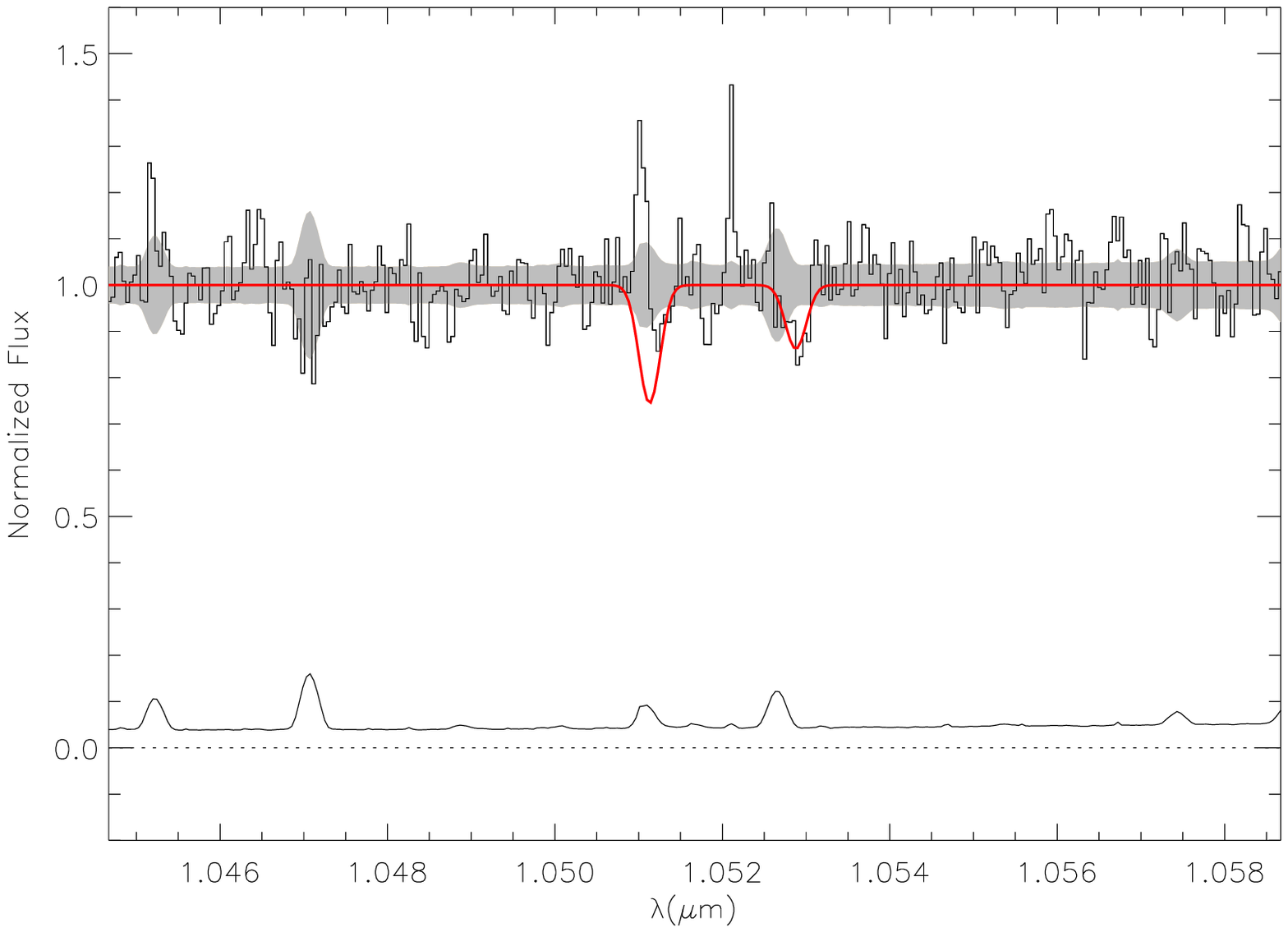}
\epsscale{1.2}
\plotone{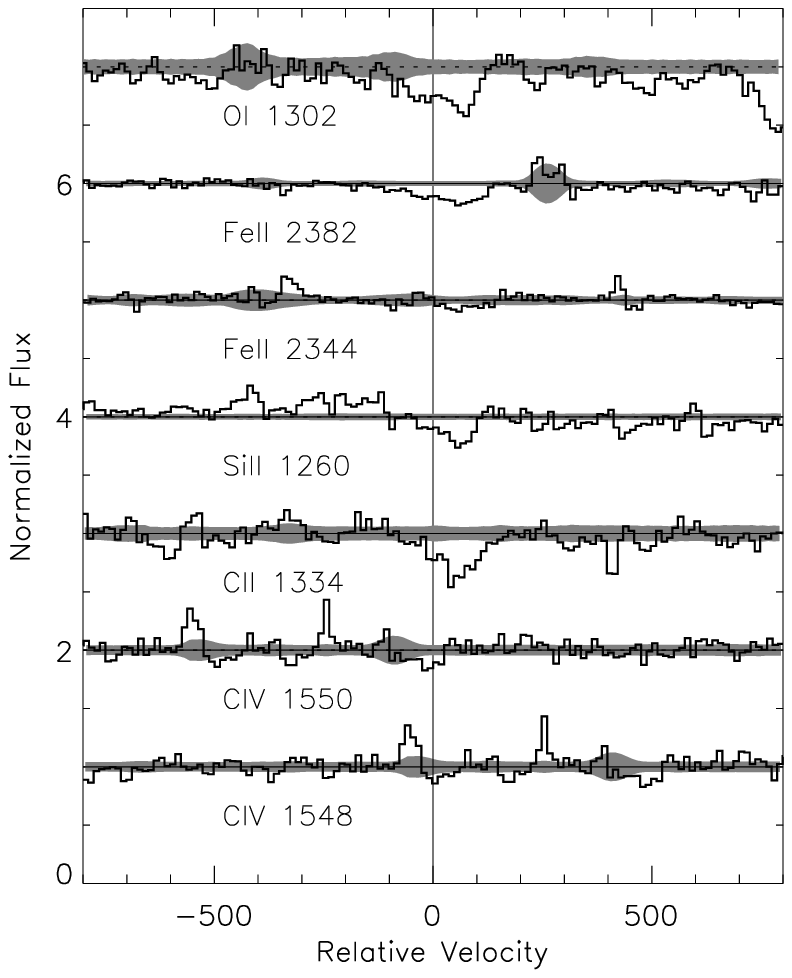}
\caption{{\em Top:} FIRE spectrum showing the \civ region of
  SDSS0818+1722 at $z=5.78990$, where both \citet{ryan_weber_civ} and
  \citet{dodorico} report tentative detection of a weak \civ system.
  Gray region shows the $\pm1\sigma$ error contours, while the solid
  black line shows a best-fit Voigt profile model to the 1550 ~\AA
  ~component alone, which does {\em not} provide a good match to the
  1548\AA ~line.  There is evidence of systematic error in the sky
  subtraction on the blue wing of the profile (the positive spike).
  {\em Bottom:} Confirmation of detected absorption in low ionization
  species at this redshift, including \oinsp, \feiinsp, \siiinsp, and \ciinsp.}
\epsscale{1.0}
\label{fig:SDSS0818_system}
\end{figure}

\citet{ryan_weber_civ} reported a possible \civ system in
SDSS0818+1722 at $z=5.7899$; this system was neither flagged in our
visual inspection of the FIRE data nor found by our automated line
search.  However, as also reported by \citet{dodorico} and
\citet{becker_cii}, we do find evidence of low ionization
absorption offset from the reported \civ redshift by $\sim 75$ \kms.
A plot of several absorption species is shown in Figure
\ref{fig:SDSS0818_system}, along with a shaded band showing the $\pm
1\sigma$ contours about the continuum.

Inspection of the error contour shows that both members of the \civ
doublet are in regions of higher noise; this is because they are both
blended with bright telluric OH emission lines at this redshift.
There is weak evidence for absorption at the 1550\AA ~component, but the
1548\AA ~component appears to suffer from systematic sky subtraction
effects.  The data for this object were obtained on a single night
(March 5, 2010) when the heliocentric velocity amplitude was 28 \kms,
i.e. nearly radial towards SDSS0818.  A revisit of this object with
seasonal offset of $\sim 1/2$ year would effectively shift the
telluric lines by one resolution element, providing an improved view
of the absorber.  

Using the 1550\AA ~line alone, we fit a Voigt profile model to check for
consistency in column density with the value reported by
\citet{ryan_weber_civ}; the result is shown in the top panel of Figure
\ref{fig:SDSS0818_system}.  Based on this component, we verified that
a $\log (N_\mciv)\approx 13.5$ \civ line represents our data quite
well.  Our fits converged to lower values of the $b$ parameter, though
these may be unphysical since the values are well below one resolution
element for FIRE.  While the velocity width is uncertain, the total
\civ column is roughly constant for any choice of $b<39$ ~\kms ~\citep[the
value measured by][]{ryan_weber_civ}.

This absorption system is intriguingly different from typical
absorption line systems at lower redshift, where the neutral phase
traced by \oinsp, \feiinsp, \siiinsp, and \cii is usually accompanied
by strong absorption from high-ionization species.  In fact the
presence of \oi absorption suggests this system may be a damped \lya
absorber or Super-Lyman-Limit system whose normal highly ionized gas
phase is substantially suppressed.

\subsubsection{Re-identifying a reported high-$z$ \civ system as \mgii}\label{sec:SDSS1030_mgii}

\begin{figure}[t]
\epsscale{1.2}
\plotone{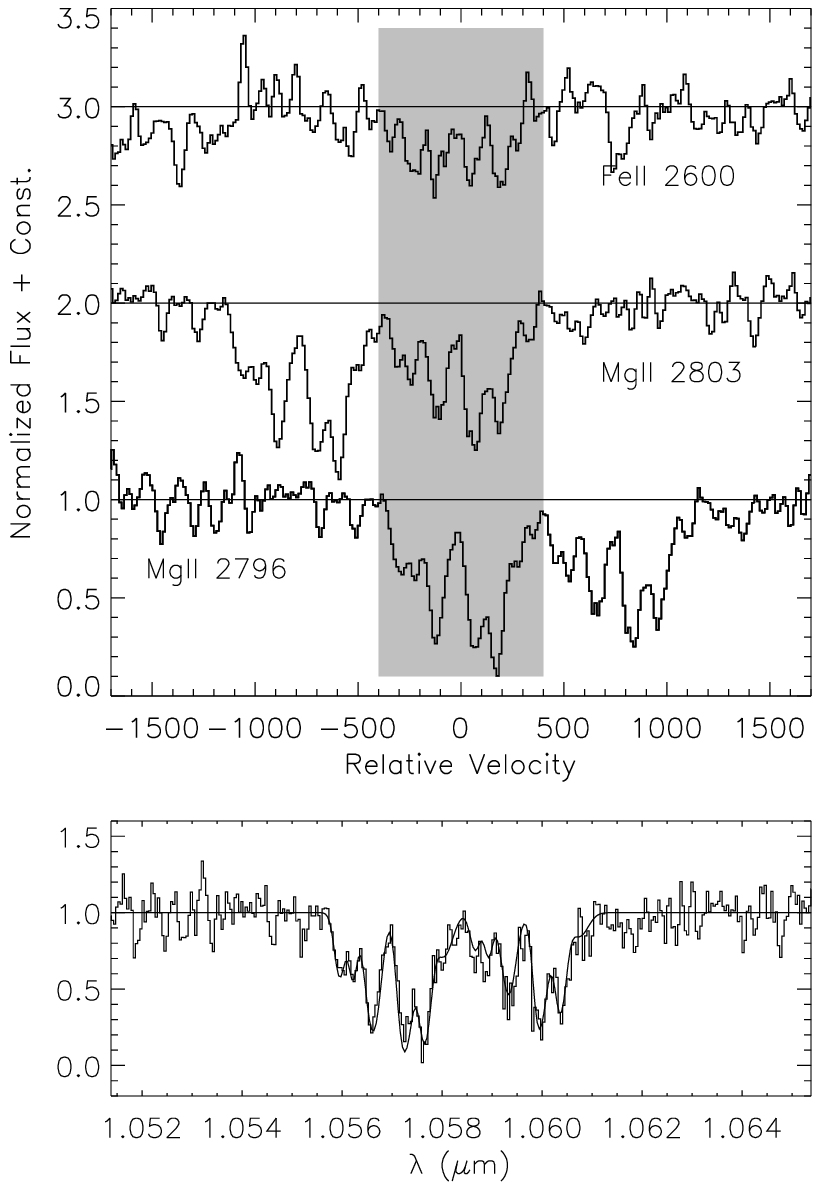}
\caption{Stacked velocity plot showing matched profiles of the \mgii
  doublet along with \feii at $z=2.780$.  This identification
  supersedes prior work which flagged this system as \civ at
  $z=5.82$.  Note the extremely broad velocity spread, which spans a
  total of $\sim 800$ \kms.  At bottom we show a Voigt profile fit to
  the \mgii complex.} \epsscale{1.0}
\label{fig:SDSS1030_mgii}
\end{figure}

Another finding with implications for the existing literature is our
improved spectrum of SDSS 1030+0524, where a strong \civ system at
$z=5.8288$ was reported by both \citet{simcoe_z6} and
\citet{ryan_weber_civ}.  This system was a marginal detection in the
ISAAC spectrum, and \citet{simcoe_z6} noted the presence of an
unidentified, blended line in his GNIRS spectrum.  In the FIRE
spectrum, we clearly identify this absorber as a \mgii system at
$z=2.780$, based on the detailed match of the \mgii 2796 and 2803\AA
~velocity profiles and the weak but detectable presence of \feii
absorption at coincident redshift.  Figure \ref{fig:SDSS1030_mgii}
shows a stacked velocity plot of three transitions from the system
along with a Voigt profile model fit for \mgiinsp.

We detect individual \mgii components\footnote{The term components
  here refers to distinct absorption features in the FIRE spectrum; at
  a resolution of $\Delta v=50$ \kms ~it is very likely that each
  spectral resolution element contains several blended and/or
  saturated intrinsic components} spanning a remarkable range of $\pm
400$ \kms, and a total equivalent width in excess of 3\AA.  This
places the system in the strongest 5\% of absorbers with $W_r>1$\AA
~found in the SDSS \citep[$>7000$ systems at $0.35<z<2.3$ in the Sloan
  DR3,][]{prochter_sdss_mgii}.  It also has a velocity width larger
than any of the 22 \mgii systems studied at high resolution in
\citet{prochter_sdss_mgii}.

The equivalent width for the $z=2.78$ absorber qualifies it as a
strong candidate Damped Lyman alpha system \citep{raoturnshek},
although the \lya line itself is inaccessible in the saturated forest.
No coincident \mgi absorption is seen, so it is uncertain whether
there is truly neutral gas present.  This leaves the physical picture
for this system uncertain.  It may represent the chance interception
of an unusually violent galactic outflow \citep{bond_superwinds}.
Alternatively if a gravitational interpretation is invoked to explain
the velocity spread, the associated potential must be somewhat large,
up to galaxy group scales, unless the system is undergoing a merger or
is otherwise out of dynamical equilibrium.  If a large galaxy or small
cluster potential is present at small enough impact parameter to be
seen in \mgii absorption (generally $\lesssim 100h^{-1}$ physical kpc
from galaxies, \citealp{Chen2010}), it is possible that this mass could
act as a gravitational lens for SDSS1030, explaining in part its large
apparent luminosity.  A full analysis of this possibility is beyond
the scope of the present paper.

\subsection{Effects on past determinations of $\Omega_\mciv$}\label{sec:changes}

The two systems discussed in sections \ref{sec:ISAAC_0818} and
\ref{sec:SDSS1030_mgii} constitute $44\%$ of the total \civ column
density (and hence 44\% of $\Omega_\mciv$) reported in the
\citet{ryan_weber_civ} survey.  The larger of these
systems---containing 38\% of the total \civnsp---is definitively removed
from the sample, requiring a downward revision of $\Omega_\mciv$ by a
factor of 1.6.  The smaller system is tentatively confirmed at the
column density reported in \citet{ryan_weber_civ}, though this
represents only a 6\% fluctuation in $\Omega_\mciv$.

Similar corrections must be applied for the results of
\citet{simcoe_z6} and \citet{becker_civ}, which on account of their
smaller sample sizes include a larger fractional contribution from the
system in SDSS1030+0524.  In the case of \citet{simcoe_z6}, the
fractional contribution is a full $56\%$, leading to a $\sim 2\times$
downward revision in the derived $\Omega_\mciv$.  Becker et al detect
{\em no} \civ systems explicitly in their NIRSPEC survey, but use the
detections of \citet{simcoe_z6} and \citet{ryan_weber_civ} to set a
lower bound on $\Omega_\mciv$; this lower bound should be revised
downward by a factor of $\sim 2$.

\section{The Universal Mass Density of \civ}\label{sec:Omega_civ}

\begin{figure*}[t]
\plotone{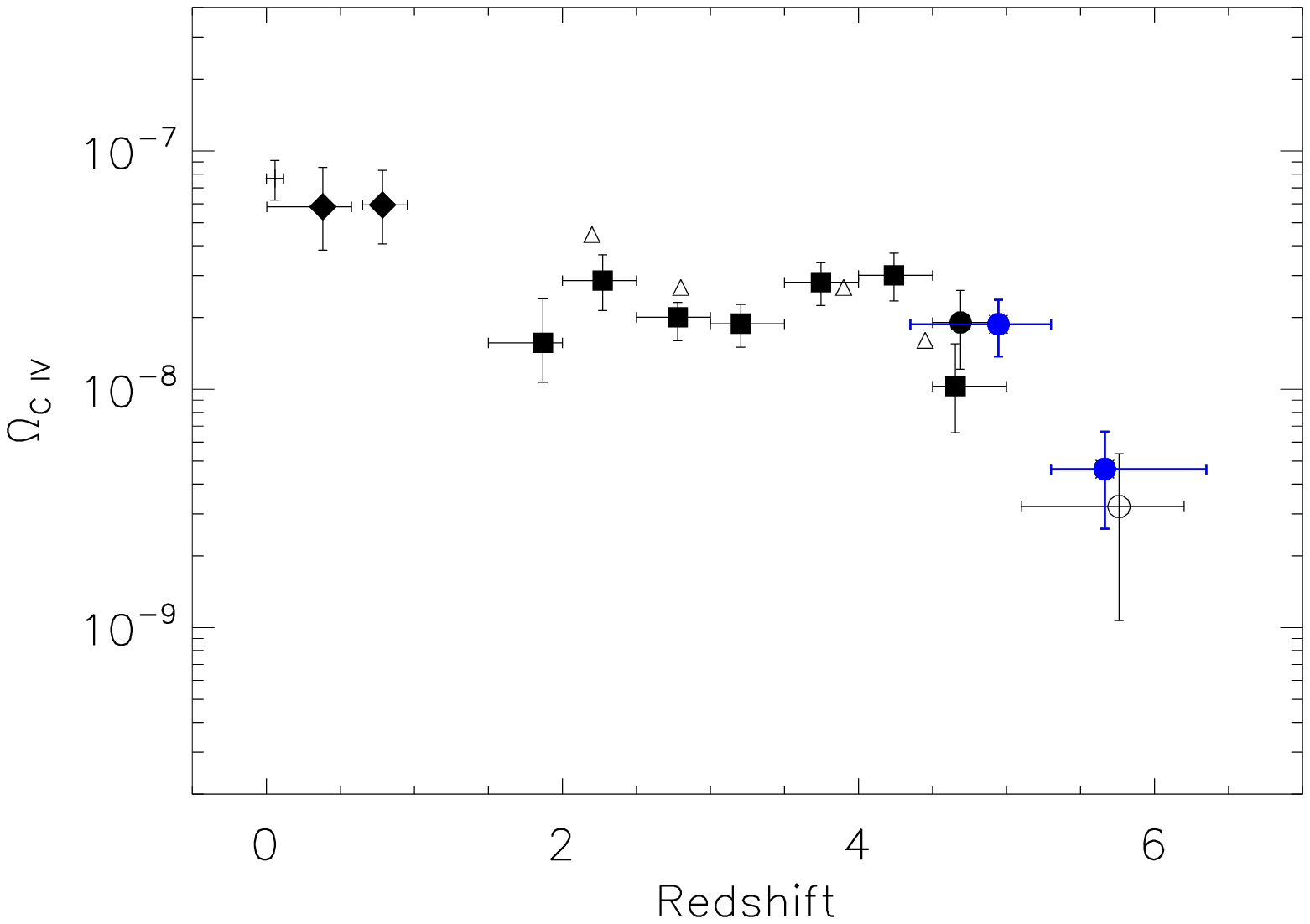}
\caption{Evolution of the mass density in \civ, expressed as
  $\Omega_\mciv$, for \civ systems with $13.4\le \log(N_\mciv)\le
  15.0$.  Data from this study are shown in red; the results of
  several prior studies are shown for comparison, where all are scaled
  to our fiducial cosmology and corrected for their varying degrees of
  completeness.  Solid squares show results from
  \citet{songaila_omegaz}, triangles are from
  \citet{songaila_new_civ}, diamonds are from \citet{cooksey_civ}, the
  cross at $z\approx 0$ is from \citet{danforthshull}, the solid black
  circle is from \citet{pettini_z5_civ}, and the open black circle is
  from \citep{ryan_weber_civ}.  Ryan-Weber's point has been adjusted
  to reflect re-identification of the putative $z=5.82$ \civ system in
  SDSS1030 as \mgiinsp.  }
\label{fig:omega_civ}
\end{figure*}

Using the \civ identifications listed in Table 2,
we constructed an estimate of $\Omega_\mciv$ using the customary form:
\begin{equation}
\Omega_\mciv = {{1}\over{\rho_c}}m_\mciv{{\sum w_i N_{\mciv,i}}\over{(c/H_0)\sum{\Delta X}}}.
\end{equation}\label{eqn:omega}
Here, $\rho_c$ represents the (current) critical density, $m_\mciv$ is
the mass of the carbon atom, and $w_i$ is the completeness correction
applied for each system, as a function of absorber $N_\mciv$ and $z$
(see Figure \ref{fig:completeness}).  $\Delta X$ represents the
comoving absorption pathlength summed over all sightlines, where for
each sightline $\Delta X = X(z_{max})-X(z_{min})$ with
\begin{equation}
X(z) = {{2}\over{3\Omega_M}}\sqrt{\Omega_M(1+z)^3+\Omega_\Lambda}.
\end{equation}
The pathlength for each sightline is listed in Table 1; the integrated
path for all FIRE spectra is $\Delta X=44.93$, of which $\Delta X =
21.92$ is at $z > 5.3$.  This is approximately a factor of $4$ larger
in path than the survey of Becker et al, and 78\% larger than the
survey of \citep{ryan_weber_civ}.  Many sightlines studied by these
authors are in the North and not accessible with FIRE.  So, to
maximize survey volume we combine separately our FIRE measurements
with (primarily NIRSPEC) measurements for northern targets to obtain a
total pathlength of $\Delta X = 62.77$, of which $\Delta X = 37.73$ is
at $z>5.3$.  Figure \ref{fig:omega_civ} shows the resulting
calculation of $\Omega_\mciv$, along with independent results at other
redshifts, gathered from the literature as listed in the figure
caption.  We have corrected the point from \citet{ryan_weber_civ} to
remove the \mgii system in SDSS1030+0524.

Several adjustments must be made to put data from different studies
together on the same plot, since each survey quotes different levels
of completeness, and in some cases different cosmological parameters
are used to calculate $\Omega_\mciv$
\citep{ryan_weber_civ,cooksey_civ}.  For older studies which use an
Einstein-deSitter cosmology ($\Omega=1$) we multiply results by
$\sqrt{\Omega_M}\approx 0.55$ to (approximately) translate to
currently favored concordance models.  A more subtle correction must
also be made for weak \civ lines not included in the sum from Equation
1 because of survey incompleteness, which varies from
study to study.

This effect may be accounted given knowledge of the column density
distribution function $f(N_\mciv) = d\mathcal{N}/dN_\mciv$ for each
survey, since $\Omega_\mciv$ may also be expressed as a weighted
integral of $f(N)$.  The power law slope of $f$ is shallower than
$N_\mciv^{-2}$ in all lower redshift surveys, so the effect of missed
systems at low $N_\mciv$ is subdominant, but could still represent a
factor of $\sim 2$ or more.  We follow the procedures outlined in
\citet{ryan_weber_civ} and \citet{cooksey_civ} to determine the
fraction of each survey's systems that would be detected in our
sample, appropriately integrating their fits to $f(N_\mciv)$ over the
range where we are complete.  We chose integration limits of
$10^{13.4}<N_\mciv<10^{15}$, since we are sensitive to many systems in
this range and our completeness corrections are robust.  The fraction
of each prior survey's systems that would be detected in the FIRE
spectra varies; the specific values we used are 0.70 for
\citet{songaila_omegaz}, 0.89 for \citet{songaila_new_civ}, 1.12 for
\citet{ryan_weber_civ}, 0.86 for \citet{pettini_z5_civ}, and 0.93 for
\citet{cooksey_civ}.  The correction for Ryan-Weber's data relies on
the unknown power law slope of $f(N)$ at $z>5$ which we assumed to
run as $N^{-1.5}$; if the slope is as steep as $N^{-1.8}$ then the
correction becomes a factor of 1.22.

After all corrections are applied, our principal result is
confirmation of a marked decline in $\Omega_\mciv$ at $z > 4.5$, as
initially reported by \citet{ryan_weber_civ} and \citet{becker_civ}.
We divided up our sample into two redshift bins with $z>5.3$ and
$z<5.3$, with the lower redshift end extending as low as $z=4.35$ and
the high redshift end to $z=6.4$.  The low redshift bin contains 14
distinct FIRE detections (where we count multi-component systems as
single detections), while the high redshift bin contains 5 FIRE
detections plus the two likely systems discussed in Section
\ref{sec:ISAAC_0818}.  Our $\Omega_\mciv$ does not include these
systems that are not formal detections with FIRE, although the
presence of low ionization absorption suggests they are probably
real.  If we treat our fits to these systems as upper limits to the
\civ column density, the amount by which $\Omega_\mciv$ could increase
is bounded at 20\% (0.08 dex).  The total result for the seven FIRE
sightlines is $\Omega_\mciv = 6.8\pm3.3 \times 10^{-9}$ at $\langle z
\rangle = 5.66$, for the column density range $13.4 <
\log(N_\mciv)<15.0$.  Similarly the value at $\langle z\rangle=4.92$
for the FIRE sample alone is $2.0\pm0.5 \times 10^{-8}$.

Given the low apparent density of \civ systems, we also re-calculated
$\Omega_\mciv$ including 6 sightlines from the literature that were
not observed by FIRE.  The additional sightlines used include
SDSS0002+2550 and SDSS1148+5251, which yielded no detections in
high-resolution (23 \kms) NIRSPEC data \citep{becker_civ}, and
SDSS0840+3549, SDSS1137+3549, SDSS1604+4244, and SDSS2054+0054, which
were observed at low resolution (185 \kms) with NIRSPEC by
\citet{ryan_weber_civ}.  This expanded set of sightlines yields just
one additional \civ system, at $z=5.738$ in SDSS1137+3529.  At this
system's $\log(N_\mciv)=14.2$, Ryan-Weber's data should be highly
complete so we apply no completeness correction to this system in our
modified sum for $\Omega_\mciv$.  Becker's spectra, while yielding no
detections, should also be highly complete over our full range of
column densities given his higher resolution.

The net effect of these largely null results is to reduce the derived
density, with our combined value for 13 sightlines measured at:

\begin{equation}
\Omega_\mciv = \left\{ {\begin{array}{l l} (1.87\pm0.50) \times 10^{-8}
    &\quad \langle z\rangle=4.95 \\ (0.46\pm0.20) \times 10^{-8} &\quad \langle z\rangle=5.66
    \\ \end{array}}\right.
\end{equation}\label{eqn:result}

The redshifts quoted above reflect the pathlength-weighted mean for
each bin, where the lower bin includes all systems from $4.35<z<5.3$
and the higher includes all systems with $5.3<z<6.4$.  While lower
than the value from the FIRE spectra alone, the different methods
agree to well within $1\sigma$.  The values in Equation 3 represent a
steady decline in $\Omega_\mciv$ by a factor of $4.1\pm2.1$ over the
$4.5<z<5.6$ interval.  This decline is measured using counts of \civ
systems drawn at different redshifts, but from the same spectra.

Our data support the contention that the \civ density is in decline at
$z>4.5$.  The exact redshift at which this downturn begins is not
strongly constrained by present data, although it appears sometime
between $z=4$ and $z=5$.  When plotted in the context of all surveys
including ones in the local universe, several interpretations are
possible regarding evolution.  Ground based studies to date have
focused on the relative constancy of $\Omega_\mciv$ from $2<z<5$,
followed by a steep drop at higher $z$.  As plotted here, an equally
plausible interpretation could be $\Omega_\mciv$ decreasing linearly
with increasing redshift from $z=0$, at constant slope until $z\sim
4.5$ after which the decline quickens.  Yet another interpretation
would hold that the \civ mass density is on relatively constant
decline over all redshifts, but is enhanced briefly at $z\sim 4$.
Coincidentally this is the redshift where the \hi
cross-section-weighted \civ ionization fraction peaks for a
\citet{haardt_cuba} prescription of the ionizing background spectrum.
At present the data do not distinguish strongly between these cases;
revisiting the $1<z<2$ and $z\sim 4$ regions with larger samples of
optical spectra should in time be able to settle the question.

\section{Discussion}

Over the past 5 years, \civ observations at $z>5$ have evolved from
samples of 2 sightlines which were strongly affected by sample
variance \citep{simcoe_z6, ryan_weber_1}, to the present sample of 13
sightlines.  Yet the number of strong detections has grown slowly,
while several sightlines nearly devoid of \civ have now been observed.
At a total detection count of 6 sure and 2 likely systems at $z>5.3$,
it is far from clear that we have sampled enough cosmic volume to
perform robust statistical analysis.

Yet the data in hand point strongly toward a lower \civ mass density
in our high redshift search region.  A critical question is to what
extent this trend reflects a change in the total carbon abundance as
opposed to a change in its ionization balance.  Rapid changes such as
this over a short timescale are more easily achieved through a
combination of radiative and ionization effects rather than chemical
feedback, but the period before $z\sim 3$ was also one of stellar mass
buildup in the universe \citep{bouwens_lfs}.  Some of the byproducts of
this process will leak into the IGM, and indeed recent estimates of
evolution in the ionization-corrected carbon abundance at $z=2-4$
suggest that the \civ mass in the IGM doubled over this period
\citep{simcoe_z4_civ}, although this represents a somewhat longer
stretch of time.

The \civ ionization balance is governed principally by photons with
energies of 3.5 Ry (\ciii to \civnsp), 4.74 Ry (\civ to \cvnsp) and $>20$ Ry
(\cv to \cvinsp).  These high-energy photons are produced predominantly
in AGN, but are also strongly attenuated by intergalactic \heii at
$z>4$, since \heii reionization likely does not occur until $z\lesssim
3.5$, and perhaps even lower redshift \citep{shull_heii_cos,
  worseck_heii}.  \civ systems---particularly the strong ones which
dominate $\Omega_\mciv$---may be preferentially present in
\heiinsp-ionized bubbles surrounding sites of galaxy formation and/or AGN
activity, where the radiation field and abundances are both enhanced.

If the evolution in the \civ abundance is driven by ionization rather
than abundance evolution, once might expect to see an increase in the
density of low ionization \ciinsp, \siiinsp, or \oi systems.  These systems
would arise in regions where abundances are high but the ionizing
field is dominated by stellar photons rather than harder AGN
radiation.

There are a few lines of evidence hinting at this interpretation for
{\em some} absorbers, though it is still too early to draw definitive
conclusions about the general population.  \citet{becker_oi} initially
reported a strong enhancement in \oi absorption toward the
high-redshift QSO SDSS1148+5251, and subsequent observations have not
yet yielded corresponding \civ absorption coincident with these \oi
lines.  More recent work by \citet{becker_cii} compiles statistics on
\cii, \oi, and \siii absorbers from HIRES and ESI spectra of a large
sample of high redshift objects.  This study suggests that the
evolution of low-ionization absorbers broadly follows the
extrapolation of trends established at lower $z$, with the highly
ionized gas phase dropping out at $z \ge 5$.  However these studies
treat the evolution of low- and high- ionization systems
independently, since examples were not available where both ionization
phases were detected in a single $z>5.5$ system.

Two of the systems studied in \citet{becker_civ} are covered by our
data on SDSS0818; one of these was first reported as a tentative \civ
detection by \citep{ryan_weber_civ} and has since also been discussed
by \citet{dodorico}.  Our data rule out strong \civ absorption in
these two systems but admit the possibility of weak lines, while we
confirm evidence of low ionization absorption.  A small handful of
systems are present in our other FIRE spectra with similar low
ionization absorption, often identified via \mgii in the $H$ band.
Analysis of these systems is outside the scope of this paper, although
it is notable that some systems contain strong \civ (and are included
in the present sample) and some do not.  At the resolution and SNR of
our FIRE spectra we do not see evidence for the emergence of a strong
``forest'' of \ciinsp, \siiinsp, or \oi lines at $z>5.5$, although higher
quality spectra will be needed to assess this possibility in detail.

Ultimately a combination of abundance and ionization evolution must be
at play, suggesting that forward modeling using numerical simulations
will be required \citep{oppenheimer, oppenheimer2, oppenheimer3,
  wiersma_metals}.  If the \civ systems reside in non-overlapping,
\heiinsp-ionized bubbles, radiative transfer effects and local enrichment
are likely to become important.  In this case the \civ systems seen at
$z>5.5$ are only ``intergalactic'' in the sense that they are selected
at random from observations of background QSOs.  In fact they are
likely to be influenced strongly by their environments in both
radiation and metal enhancement, and may reside quite close to the
early galaxies responsible for reionizing the universe.

\section{Conclusions}

Using the newly commissioned FIRE spectrograph, we have observed a
sample of seven QSOs at $z>5.5$ to derive improved constraints on the
\civ abundance at early times.  Using both manual and automated
searches, we identified a sample of 19 lines in our FIRE spectra with
\civ column densities in the range $13.2 \le \log(N_\mciv)\le14.6$,
and redshifts ranging from $z=4.61$ to $5.91$.  Monte Carlo tests
indicate that our line sample is highly complete at $N_\mciv >
10^{14}$~\pcmsq; it is 50\% complete near $10^{13.7}$~\pcmsq over most
survey areas not strongly affected by telluric absorption.

Our conclusions may be summarized as follows:

\begin{enumerate}
\item{Our spectra show evidence of a decline in the \civ abundance at
  $z>5$, consistent with the findings of previous studies not strongly
  affected by cosmic variance.  Expressed in terms of the closure
  density, we find $\Omega_\mciv=4.6\pm2.0\times10^{-9}$ at $\langle z
  \rangle=5.66$, a four-fold decline relative to what is observed at
  $\langle z\rangle=4.95$ and lower redshifts.}
\item{A strong system in the spectrum of SDSS1030+0524 previously
  identified in the literature as \civ at $z=5.82$ is re-identified as
  \mgii at $z=2.78$.  This system comprised a significant portion of
  the total \civ mass measured in all prior surveys, so these
  literature results must accordingly be revised downward.}
\item{We present new limits on the \civ column density for several
  systems previously identified based on low-ionization lines and/or
  weak \civnsp.  The emergence of these systems may reflect an overall
  change in the ionization balance of absorption systems at $z>5$,
  where high-ionization species common at low redshift are less
  prevalent.  Larger statistical samples of these systems could yield
  interesting model constraints on the evolution of the ionizing
  background versus metallicity fields}
\end{enumerate}

Finally, we caution that as the \civ density drops, even though the
number of well-observed sightlines grows our results may still be
subject to fluctuations from shot noise or sample variance (only $5-7$
systems are now known at $z>5.5$).  The present work summarizes just
the first year of an ongoing program to observe high-redshift QSOs.
Planned observations should cover the complete catalog of bright,
southern-accessible targets, while ongoing and planned surveys such as
UKIDSS \citep{ukidss}, SkyMapper \citep{skymapper}, and VISTA
\citep{vista} should substantially increase the number of known
high-$z$ quasars in the Southern sky.

\bigskip

{\it Facilities:} \facility{Magellan:Baade (FIRE)}

\acknowledgements

It is a pleasure to thank the many engineers and staff at MIT and
Magellan who contributed to FIRE's successful construction and
commissioning.  We gratefully acknowledge support for FIRE's
construction through the NSF under MRI grant AST-0619490, and also by
the MIT Department of Physics and Curtis Marble.  Support for science
observations and analysis was provided by NSF grant AST-0908920, and
also by the Alfred P. Sloan Foundation.  RAS enthusiastically
recognizes generous lumbar support from the Adam J. Burgasser Endowed
Chair in Astrophysics.  AJB acknowledges financial support from the
Chris and Warren Hellman Fellowship Program.

\bibliography{z6civ}

\bigskip

\begin{deluxetable*}{lccccccc}
\tablecaption{\civ Detections}
\tablehead{
\colhead{Object} &
\colhead{$z_{abs}$} &
\colhead{$W_r(1548)$\tablenotemark{1}} &
\colhead{$W_r(1550)$\tablenotemark{1}} &
\colhead{$N_\mciv(AODM\tablenotemark{2})$} &
\colhead{$N_\mciv(VPFIT)$\tablenotemark{3}} &
\colhead{$b_\mciv(VPFIT)$\tablenotemark{4}} &
\colhead{$\log_{10}(w_i)$\tablenotemark{5}}}
\startdata
SDSS0818+1722  & 4.69193 & $115\pm23$   & $63\pm 32$  & $13.3\pm0.20$  & $13.43\pm0.09$ & $26\pm13$           & 0.49 \\
\nodata        & 4.72586 & $462\pm25$   & $290\pm17$  & $>14.1\pm0.03$  & $13.96\pm0.04$ & $43\pm8$            & 0.05 \\
\nodata        & 4.72737 & \nodata      &  \nodata    &\nodata          & $13.67\pm0.15$ & $10$                & 0.19 \\
\nodata        & 5.78917\tablenotemark{6}& $37\pm 20$ & $54$  & \nodata & $\le 13.1$ & \nodata                & \nodata \\
\nodata        & 5.87689\tablenotemark{6}& $47\pm13$  & $33\pm13$ & $13.3\pm0.08$ & $\le 13.41$ & $6.7\pm8.0$ & \nodata \\
\hline
CFHQS1509-1749 & 4.61078 & $26\pm7$     & $11\pm8$    & $12.8\pm0.16$  & $13.23\pm0.28$ & $5.0\pm6.0$         & 0.87 \\
\nodata        & 4.64143 & $71\pm10$    & $60\pm10$   & $13.4\pm0.10$   & $13.44\pm0.04$ & $48.8\pm7.0$        & 0.43 \\
\nodata        & 4.66598 & $89\pm14$    & $35\pm11$   & $13.3\pm0.10$    & $13.41\pm0.05$ & $40.3\pm7.7$        & 0.50 \\
\nodata        & 4.81552 & $308\pm22$   & $206\pm 19$ & $>14.0\pm 0.10$   & $14.01\pm0.04$ & $60.0\pm5.2$        & 0.06 \\
\nodata        & 4.82492 & \nodata      & $33\pm12$   & $13.2\pm 0.20$   & $13.45\pm0.09$ & $26.0\pm10.0$       & 0.63 \\
\nodata        & 5.91578 & $331\pm39$   & $183\pm25$  & $>14.1\pm0.07$  & $14.02\pm0.03$ & $48.8\pm5.5$        & 0.03 \\
\hline
SDSS1030+0524  & 4.94814 & $109\pm27$   & $237\pm27$  & $13.9\pm0.07$  & $13.76\pm0.11$ & $33.8\pm23.1$       & 0.57 \\
\nodata        & 5.51694 & $331\pm31$   & $256\pm23$  & $>14.0\pm0.04$  & $14.02\pm0.04$ & $72.9\pm12.0$       & 0.07 \\
\nodata        & 5.72438 & $698\pm17$   & $482\pm18$  & $>14.6\pm0.03$   & $14.59\pm0.04$ & $50.3\pm3.0$        & 0.04 \\
\nodata        & 5.74399 & $273\pm15$   & $172\pm13$  & $13.87\pm0.05$  & $14.00\pm0.04$ & $40.2\pm5.7$        & 0.05 \\
\hline
SDSS0836+0054  & 4.99623 & $163\pm17$   & $73\pm17$   & $13.6\pm0.06$   & $13.70\pm0.05$ & $71.1\pm11.5$       & 0.63 \\
\hline
SDSS1411+1217  & 5.24988 & $734\pm26$   & $321\pm22$  & $>14.3\pm0.04$   & $14.14\pm0.07$ & $30.0$\tablenotemark{7} & 0.03\\
\nodata        & 5.24788 & \nodata      & \nodata     & \nodata         & $13.66\pm0.10$ & $30.0$\tablenotemark{7} & 0.32 \\ 
\nodata        & 5.25151 & \nodata      & \nodata     & \nodata         & $13.95\pm0.07$ & $30.0$\tablenotemark{7} & 0.11 \\ 
\nodata        & 5.786\tablenotemark{8} & \nodata     & \nodata         & \nodata & \nodata & \nodata            & \\
\hline
SDSS1306+0356  & 4.61499 & $320\pm13$   & $184\pm14$  & $13.9\pm0.03$   & $14.09\pm0.15$  & $64.4\pm4.7$          & 0.02 \\
\nodata        & 4.66825 & $292\pm31$   & $198\pm22$  & $>14.0\pm0.05$   & $14.07\pm0.03$ & $58.1\pm5.8$          & 0.02 \\
\nodata        & 4.86591 & $1744\pm50$  & $901\pm41$  & $>14.7\pm0.03$   & $14.80\pm0.01$ & $80.6\pm32.3 $                         & 0.05 \\
\nodata        & 4.87965 & $514\pm32$   & $354\pm30$  & $>14.4\pm0.10$    & $14.30\pm0.20$ & $55.6\pm19.4$         & 0.06 \\ 
\hline
ULAS1319+0905 & 5.57415   & $392\pm44$   & $196\pm51$  & $>14.1\pm 0.10$   & $14.14\pm0.04$ & $65.5\pm9.1$         & 0.04 \\
\nodata       & 5.26490   & $133\pm33$   & $118\pm27$  & $13.6\pm 0.15$  & $13.66\pm0.10$ & $62.9\pm21.7$        & 0.32
\enddata

\tablenotetext{1}{Rest frame equivlent width (m\AA).}
\tablenotetext{2}{Column density  (\pcmsq) determined via the apparent optical depth method \citep{AODM}.  The AODM is formally a lower bound on the column density for saturated doublets, so systems with $N_\mciv>10^{14}$ \pcmsq ~are listed as limits in the table.}
\tablenotetext{3}{Voigt profile column density (\pcmsq).}
\tablenotetext{4}{Voigt profile doppler parameter (\kms).}
\tablenotetext{5}{Completeness correction (dex)}
\tablenotetext{6}{This system is not a statistically significant \civ system in our spectrum, but we confirm the presence of \siii and other neutral atoms as reported by \citet{dodorico} and \citet{ryan_weber_civ}.  It is {\em not} included in the $\Omega_\mciv$ calculation.}
\tablenotetext{7}{This system is
  blended such that the uncertainty in $b$ was quite high; this has
  minimal effect on the total column density.} 
\tablenotetext{8}{Intrinsic BAL.} 

\end{deluxetable*}

\end{document}